# Primordial Non-Gaussian Signatures in the Sky

*Alejandro Gangui*

*SISSA - International School for Advanced Studies. Trieste*

*The presence of non-Gaussian features in the CMB radiation maps represents one of the most long-awaited clues in the search for the actual structure of the primordial radiation. These features could shed some light on the non trivial task of distinguishing the real source of the primeval perturbations leading to large scale structure. In the present paper we briefly review recent work towards finding analytical estimates of the three- and four-point correlation functions and of their zero-lag limits, namely, the skewness and kurtosis, respectively.*

A couple of years ago, anisotropies in the CMB radiation were finally detected by the COBE satellite. This fact and the need to disentangle the relevant information out of the primeval radiation maps boosted a large body of literature dealing with the non trivial task of finding the correct statistics. Two are the current favourite theories for the origin of the primordial perturbations, namely, seeds due to topological defects produced during a GUT phase transition and quantum fluctuations produced during an early inflationary era. Present observational data have not yet been able to rule out one of them. A clear-cut discriminating tool between these two scenarios still proves difficult to find. It is a commonly accepted fact that Inflation produces fluctuations that are gaussian. There is a full closet of possible (and attractive) models with very different motivations but common aims: render the cosmological standard model free of its (few) blemishes while at the same time create the conditions for the generation of LSS. These (too) many models not always coincide in their predictions (although all of them should do a god job concerning, say, the flatness, horizon and unwanted relics problems). The predicted value for the spectral index is one point of contrast among them (but usually the results of data analyses come with errors large enough to make nearly all of them "marginally consistent" and nobody questions too much this aspect [1]. Another feature is whether they leave room for the production of tensor perturbations (gravitational waves). Of course this includes another free parameter in the scenario, because "which is the amount of the tensor contribution to the detected quadrupole", nobody is yet able to tell (see Lidsey's contribution to these proceedings). Other approaches face the problem differently. Large scale surveys (anisotropies: COBE,FIRS,TEN,SP. . .; galaxy surveys: APM,CfA,IRAS; peculiar velocities: POTENT) provide data today that is the result of primordial perturbations in the inflaton $\sim 60 - 50$ e-foldings before the end of the inflationary era and therefore should serve us well for reconstructing at least a small piece of $V(\phi)$ (and may be the whole of it some day-if at all possible) (see Kolb's contribution). What about defects? Let's concentrate on cosmic strings for definiteness. They certainly provide a very appealing scenario. Unlike Inflation, strings were thought of as being sources for non-gaussianness in the CMB. Possessing a good deal of elegance, this theory is able to account for a host of observational facts just by choosing adequately its single free parameter $G\mu = (v/m_P)^2$ (where $\mu$ is the effective mass per unit length of the wiggly string and $G$ is Newton's constant). This mass-scale ratio is in fact a very attractive feature and even models of inflation have been proposed [2] where observational predictions are related to similar mass scale relations. Cosmic strings may be incorporated into every fashionable cosmological model. Their presence helps in accounting for the formation of large scale filaments and sheets, galaxy formation at epochs $z \sim 2 - 3$, and galactic magnetic fields. They also generate peculiar velocities on large scales, and are consistent with the amplitude, spectral index and the statistics of the measured CMB anisotropies (see Shellard's contribution). One possible way to discriminate between these two theories may be found in the structure of the CMB anisotropies. Several methods and statistics have been proposed so far, being their principal aim to provide a tool that would enable us to find a "primordial non-gaussian signature in the sky". What follows is a short account of some recent work on higher order temperature correlation functions suitable for both scenarios. We begin with the description of the inflation-generated three-point correlation function and its possible collapsed cases, including the skewness. Then, we sketch briefly some steps towards the computation of the four-point function and the kurtosis from cosmic strings. As a first step we will define the connected two- and three-point correlation functions of the CMB temperature as measured by a given observer, i.e. on a single microwave sky. The angular two-point correlation function $C_2(\mathbf{x}; \alpha)$ measured by an observer placed in $\mathbf{x}$ is the average product of temperature fluctuations in two directions $\hat{\gamma}_1$ and $\hat{\gamma}_2$ whose angular separation is $\alpha$; this can be written as

$$C_2(\mathbf{x}; \alpha) = \int \frac{d\Omega_{\hat{\gamma}_1}}{4\pi} \int \frac{d\Omega_{\hat{\gamma}_2}}{2\pi} \delta(\hat{\gamma}_1 \cdot \hat{\gamma}_2 - \cos\alpha) \frac{\Delta T}{T}(\mathbf{x}; \hat{\gamma}_1) \frac{\Delta T}{T}(\mathbf{x}; \hat{\gamma}_2) . \quad (1)$$

As well known, in the limit $\alpha \to 0$ one recovers the CMB variance. Expanding the temperature fluctuation in spherical harmonics $\frac{\Delta T}{T}(\mathbf{x}; \hat{\gamma}) = \sum_{\ell=1}^{\infty} \sum_{m=-\ell}^{\ell} a_\ell^m(\mathbf{x}) \mathcal{W}_\ell Y_\ell^m(\hat{\gamma})$, we easily arrive at the expression $C_2(\mathbf{x}; \alpha) = \frac{1}{4\pi} \sum_\ell P_\ell(\cos\alpha) Q_\ell^2(\mathbf{x}) \mathcal{W}_\ell^2$ where $Q_\ell^2 = \sum_{m=-\ell}^{\ell} |a_\ell^m|^2$. The analogous expression for the angular three-point correlation function is obtained by taking the average product of temperature fluctuations in three directions $\hat{\gamma}_1$, $\hat{\gamma}_2$ and $\hat{\gamma}_3$ with fixed angular separations $\alpha$ (between $\hat{\gamma}_1$ and $\hat{\gamma}_2$), $\beta$ (between $\hat{\gamma}_2$ and $\hat{\gamma}_3$) and $\gamma$ (between $\hat{\gamma}_1$ and $\hat{\gamma}_3$); these angles have to satisfy the obvious inequalities $|\alpha - \gamma| \leq \beta \leq \alpha + \gamma$. One has

$$C_3(\mathbf{x}; \alpha, \beta, \gamma) = N(\alpha, \beta, \gamma) \int \frac{d\Omega_{\hat{\gamma}_1}}{4\pi} \int \frac{d\Omega_{\hat{\gamma}_2}}{2\pi} \int \frac{d\Omega_{\hat{\gamma}_3}}{2} \delta(\hat{\gamma}_1 \cdot \hat{\gamma}_2 - \cos\alpha)$$
$$\times \delta(\hat{\gamma}_2 \cdot \hat{\gamma}_3 - \cos\beta) \delta(\hat{\gamma}_1 \cdot \hat{\gamma}_3 - \cos\gamma) \frac{\Delta T}{T}(\mathbf{x}; \hat{\gamma}_1) \frac{\Delta T}{T}(\mathbf{x}; \hat{\gamma}_2) \frac{\Delta T}{T}(\mathbf{x}; \hat{\gamma}_3) , \quad (2)$$

where $N(\alpha, \beta, \gamma) \equiv \sqrt{1 - \cos^2\alpha - \cos^2\beta - \cos^2\gamma + 2\cos\alpha\cos\beta\cos\gamma}$. Setting $\alpha = \beta = \gamma = 0$ in these general expressions one obtains the CMB *skewness* $C_3(\mathbf{x}) = \int \frac{d\Omega_{\hat{\gamma}}}{4\pi} [\frac{\Delta T}{T}(\mathbf{x}; \hat{\gamma})]^3$. Also useful are the *equilateral* three-point correlation function and the *collapsed* one, corresponding to the choices $\alpha = \beta = \gamma$, and $\alpha = \gamma$, $\beta = 0$, respectively. Following a standard procedure we can rewrite the three-point function in the form

$$C_3(\mathbf{x};\alpha,\beta,\gamma) = N(\alpha,\beta,\gamma)\frac{\pi}{2} \sum_{\ell_1,\ell_2,\ell_3} \sum_{m_1,m_2,m_3} a_{\ell_1}^{m_1} a_{\ell_2}^{m_2} a_{\ell_3}^{m_3\,*} \mathcal{W}_{\ell_1} \mathcal{W}_{\ell_2} \mathcal{W}_{\ell_3}$$

$$\times \sum_{j,k,\ell} \sum_{m_j,m_k,m_\ell} P_j(\cos\alpha) P_k(\cos\beta) P_\ell(\cos\gamma) \mathcal{H}_{j\ell\ell_1}^{m_j m_\ell m_1} \mathcal{H}_{kj\ell_2}^{m_k m_j m_2} \mathcal{H}_{k\ell\ell_3}^{m_k m_\ell m_3} \quad (3)$$

where $\mathcal{H}_{\ell_1\ell_2\ell_3}^{m_1 m_2 m_3} \equiv \int d\Omega_{\hat{\gamma}} Y_{\ell_1}^{m_1\,*}(\hat{\gamma}) Y_{\ell_2}^{m_1}(\hat{\gamma}) Y_{\ell_3}^{m_3}(\hat{\gamma})$ can be expressed in terms of Clebsch-Gordan coefficients. Analogous expressions may be found for the four-point function, but their form is slightly more involved and will not be presented here. So far our expressions have been kept completely general, they would apply to whatever source of temperature fluctuations in the sky, through suitable (usually statistical) relations for the product of three multipole coefficients $a_\ell^m$ appearing in Eq.(3), and to whatever angular scale, through the specific choice of window functions $\mathcal{W}_\ell$. In what follows we shall only briefly sketch some results where we deal with large angular scale anisotropies originated from primary perturbations in the gravitational potential $\Phi$ on the last scattering surface via the Sachs-Wolfe effect. To obtain definite predictions for the statistics described above, one needs to exploit the random nature of the multipole coefficients $a_\ell^m$. In our case, these coefficients should be considered as zero-mean non-Gaussian random variables whose statistics should in principle be deduced from that of the gravitational potential. We here compute analytically the theoretical ensemble expectation values in order to obtain the mean two- and three-point functions. These expectation values are of course observer- i.e. $\mathbf{x}$-independent and can only depend upon the needed number of angular separations. In the frame of the stochastic approach to inflation (see Starobinskii's contribution), the calculations reported in Ref.[3] lead to general expressions for the angular spectrum, $\langle a_{\ell_1}^{m_1} a_{\ell_2}^{m_2\,*} \rangle = \delta_{\ell_1\ell_2} \delta_{m_1 m_2} \mathcal{Q}^2 \mathcal{C}_{\ell_1}/5$, and the angular bispectrum

$$\langle a_{\ell_1}^{m_1} a_{\ell_2}^{m_2} a_{\ell_3}^{m_3\,*} \rangle = \frac{3\mathcal{Q}^4}{25} \Phi_3 \left[ \mathcal{C}_{\ell_1}\mathcal{C}_{\ell_2} + \mathcal{C}_{\ell_2}\mathcal{C}_{\ell_3} + \mathcal{C}_{\ell_3}\mathcal{C}_{\ell_1} \right] \mathcal{H}_{\ell_3\ell_1\ell_2}^{m_3 m_1 m_2} \quad (4)$$

where $\Phi_3$ is a model-dependent coefficient, $\mathcal{Q} = \langle Q_2^2 \rangle^{1/2}$ is the *rms* quadrupole and $\mathcal{C}_\ell$ is defined by $\langle Q_\ell^2 \rangle \equiv (2\ell+1)\mathcal{Q}^2 \mathcal{C}_\ell/5$. Replacing (4) into (3) we obtain the general form of the mean three-point correlation function. The CMB mean skewness $\langle C_3(0) \rangle$ immediately follows from the above equations. The stochastic analysis naturally takes into account all the multiplicative effects in the inflaton dynamics that are responsible for the non-Gaussian features. Extra-contributions to the three-point function of the gravitational potential also arise as a consequence of the non-linear relation between $\Phi$ and the perturbation in the inflaton field $\delta\phi$. When all primordial second-order effects are taken into account we get for the "dimensionless" skewness $\mathcal{S} \equiv \langle C_3(0) \rangle / \langle C_2(0) \rangle^{3/2}$ the expression $\mathcal{S} \simeq (\sqrt{45\pi}/32\pi^2)\mathcal{Q} \left[ X_{60}^2 - 4m_P X_{60}' \right]$ where we denoted with $X_{60}$ the value of the steepness of the inlaton potential $X(\phi) = m_P V'(\phi)/V(\phi)$ evaluated at $\phi_{60}$ (60 e-foldings before the end). One may now apply the above formulae to specific models. The actual values found for all fashionable single-field models is too small to be distinguished from the theoretical uncertainties that dominate the large scale observations. Simply, the multiplicative effects responsible for the non-gaussian features have not enough time for building up a specific signature able to emerge from behind the cosmic variance reigning at COBE scales.

Strings may also leave their imprint on the CMB in several different ways. The best studied mechanism for producing temperature fluctuations is the Kaiser-Stebbins effect. According to this effect, moving long strings present between the time of recombination $t_{rec}$ and today produce (due to their deficit angle) discontinuities in the CMB temperature between photons reaching the observer through opposite sides of the string. Other mechanisms for producing CMB fluctuations by cosmic strings are i) potential fluctuations on the last scattering surface produced by long strings and loops present between $t_{eq}$ and $t_{rec}$. ii) Doppler effect by moving long strings present on the LSS. All these effects should be superposed in order to calculate the full spectrum. However, for large scales the Kaiser-Stebbins effect dominates and therefore we will concentrate solely on this in what follows. We will follow the multiple impulse approach, which although simplified, has the advantage of being analytical. The magnitude of the discontinuity is proportional not only to the deficit angle but also to the string velocity $v_s$ and depends on the relative orientation between the unit vector along the string $\hat{s}$ and the unit photon wave-vector $\hat{k}$. It is given by $\Delta T/T = \pm 4\pi G\mu v_s \gamma_s \hat{k} \cdot (\hat{v}_s \times \hat{s})$, where $\gamma_s$ is the relativistic Lorentz factor and the sign changes when the string is crossed. Long strings within each horizon have random velocities, positions and orientations. In this framework, the effect of the string network on a photon beam is just the linear superposition of the individual effects, taking into account compensation. This leads to the fact that *kicks* suffered by photon beams separated by an angular scale larger than the horizon size at a certain time-step are uncorrelated, whereas afterwards when the horizon size gets larger (and when the beam angular separation *fits* in the horizon angular scale) they will be influenced by the same strings and therefore these perturbations become correlated. This procedure may be fully worked out and the general N-point correlation function may be derived. However, the superimposed kernels of the distribution turn out to be symmetric with respect to positive and negative perturbations and therefore no mean value for the skewness arises. On the other hand the four-point function is easily found [3] and a value for the kurtosis parameter may be predicted. Nonetheless, as we could expect, this primordial signal gets blurred when the number of seeds is increased. Future work will certainly broaden our way towards a cleanest picture in this interesting subject.

**Acknowledgements:** The author is grateful to P. Dunsby, L. Kofman, D. Lyth and J. Mather for concise and enlightening discussions. He also thanks F. Occhionero and colleagues for the organization of an instructive and charming workshop. This work was supported by the Italian MURST.